\documentclass[twocolumn,prb]{revtex4}
\usepackage[T1]{fontenc}
\usepackage[utf8]{inputenc}
\usepackage{amsmath,amssymb}
\usepackage{color}
\usepackage{graphicx}
\usepackage{amssymb}
\begin{document}
\title{Charge redistribution and interlayer coupling in twisted bilayer graphene under electric fields}
\author{E. Su\'{a}rez Morell, P. Vargas}
\affiliation{Departamento de F\'{i}sica, Universidad T\'{e}cnica
Federico Santa Mar\'{i}a, Casilla 110-V, Valpara\'{i}so, Chile}
\author{ L. Chico and L. Brey  }
\affiliation{Departamento de Teor\'{\i}a y Simulaci\'on de Materiales, Instituto de Ciencia de Materiales de Madrid, CSIC, 28049 Cantoblanco, Spain}

\begin{abstract}
We investigate the electronic density redistribution of rotated bilayer graphene under a perpendicular electric field, showing that the layers are actually coupled even for large angles. This layer-layer coupling is evidenced by the charge transfer on these structures as a function of the external voltage. We find an inhomogeneous excess charge distribution that is related to the moir\'e patterns for small angles, but that persists for larger angles where the carriers' velocity is equal to that of single layer graphene. Our results show that rotated bilayer systems are coupled for all rotation angles.
\end{abstract}

\maketitle

\section{Introduction}
The electronic properties of graphene can be modified by piling up a few layers, thus changing the behavior of the charge carriers due to the interlayer coupling. One of the main reasons behind the interest in bilayer graphene (BLG) is precisely that its low-energy properties are different from those of monolayer graphene.
Bilayer graphene with Bernal  (or AB) stacking has massive chiral fermions, and a gap can be opened by applying an electric field.\cite{McCann_2006} In contradistinction, the dispersion relation of BLG with direct (or AA) stacking is linear, and remains gapless with an applied gate voltage. Even though Bernal stacking is the lowest energy arrangement for bulk graphite, in few-layer graphene other stack orderings are possible: indeed, bilayer graphene with AA stacking has been observed,\cite{Liu_2009} as well as twisted bilayer graphene,\cite{Hass_2008,Ohta2006,Reina2009} 
which consists of two adjacent graphene layers where one of the layers is rotated with respect to the other, resulting in larger unit cells with intriguing electronic properties.\cite{Lopes_2007,EricFlatBands,Trambly_2010,Bistritzer2011}

As a matter of fact, there is an ongoing controversy on the electronic characteristics of rotated bilayer graphene. 
Several experiments on rotated few-layer graphene grown on SiC show an electronic behavior similar to that of single-layer graphene, with the same carriers' velocity as 
that of an isolated graphene monolayer; for this reason, these systems have been considered as composed of uncoupled graphene sheets.
\cite{Hass_2008,Varchon_2008,Song_2010,Hicks_2011} However, experimental results in twisted graphene bilayers fabricated by chemical vapor deposition indicate that twisted graphene bilayers may be coupled, especially for small rotation angles, as evidenced by  the appearance of low-energy van Hove singularities\cite{Li_2010} 
and the measured renormalization of the carrier velocity.\cite{Luican_2011} 

From the theoretical viewpoint, it is established that twisted BLG with a relative rotation angle (RRA) greater than 10$^{\rm o}$ presents a linear dispersion relation with the same velocity as monolayer graphene.\cite{Lopes_2007,Shallcross_2008,Trambly_2010} For RRA  between 1$^{\rm o}$ and 10$^{\rm o}$, the carrier velocity diminishes, 
as evidenced by continuum\cite{Lopes_2007} and combined tight-binding/first-principles calculations.\cite{Trambly_2010} For smaller angles, around 1$^{\rm o}$,  flat bands appear close to the  Fermi energy.\cite{EricFlatBands,Trambly_2010} 
However, for certain rotation angles where theoretical calculations predict a renormalization of the carriers' velocity, some experimental results have shown a behavior like that of monolayer graphene \cite{Song_2010, Hicks_2011} while others do obtain the predicted renormalized velocity for small RRA.\cite{Luican_2011} 
In principle, 
the key to understand this disagreement is to elucidate the coupling strength in adjacent graphene layers.\cite{MacDonald_2011}

In this paper we propose a way to assess the interlayer interaction in twisted bilayer graphene 
by exploring the spatial distribution of the electronic density. 
We show  that, in spite of the linear dispersion relation and the Fermi velocity of these systems, the layers can be actually coupled even for large rotation angles. The layer-layer coupling is evidenced by the charge transfer between layers under an applied electric field. In addition, we find an inhomogeneous excess charge 
for small angles,  persisting for larger angles where the carriers' velocity is equal to that of single layer graphene. We have found that the average excess charge density on a layer of a twisted bilayer graphene, due to an applied electric field, depends on the RRA. The charge transfer is larger the smaller the angle but even for RRA above 20$^{\rm o}$ there is a significant charge transfer.
The charge has an inhomogeneous distribution over the twisted unit cell, with a relative weight on sites A and B also depending on the angle.

This article is organized as follows. In Section \ref{sec:geo_mod} we describe the geometry of the commensurate graphene bilayers and the model employed to calculate the electronic properties. Section \ref{sec:results} is devoted to explain and discuss our results. We conclude in Sec. \ref{sec:concl}, where we summarize our main findings.

\section{Geometry and Model}
 \label{sec:geo_mod}
\subsection{Geometry}

The system studied consists on two graphene layers which are rotated from a Bernal stacking configuration, subject to a potential difference. We consider the top layer to be at a positive potential $+V$ and the bottom layer at $-V$. 
This potential difference produces an electronic charge transfer from the bottom to the top layer.  Consequently, an electronic excess density $n$ builds up on the top layer.   We will show that the excess charge depends on the interlayer coupling. 

In order to build a commensurate unit cell we follow a procedure similar to that of Campanera {\it et al.}, finding coincidence lattice points in the BLG crystal.\cite{Campanera_2007}  We start with two graphene layers with Bernal (AB) stacking. For this arrangement there are two non-equivalent sites within a layer, namely, the A
 site, where an atom of the upper layer lies directly on top of another atom of the lower layer, and the B 
 site, where the atom is exactly at the center of the hexagon of the lower layer. We have chosen a B site as our rotation center. We do a clockwise commensurable rotation from a vector $\vec{r}=m \vec{a}_{1} + n \vec{a}_{2} $ to  $\vec{t}_{1}=n \vec{a}_{1} + m \vec{a}_{2}$, where $\vec{a}_{1}=(-1/2,\sqrt{3}/2)a_{0}$ and  $\vec{a}_{2}=(1/2,\sqrt{3}/2)a_{0}$ are the graphene bilayer lattice vectors; \textit{n},\textit{m} are integers, and $a_{0}=2.46\,$\AA\  is the lattice constant. The unit cell vectors can be chosen as $ \vec{t}_{1}=n \vec{a}_{1} + m \vec{a}_{2} $ and $\vec{t}_{2}= -m \vec{a}_{1} + (n+m) \vec{a}_{2}$. All magnitudes needed, like the relative rotational angle (RRA), the number of atoms in the unit cell and the reciprocal lattice vectors, can be expressed as functions of $n,m$ and $a_{0}$. For instance, the RRA is given by
 $ \cos \theta= (n^{2}+ 4mn+m^{2})/2(n^{2}+ mn+m^{2})$. 
Henceforth, we label a commensurate unit cell by the indices $(n,m)$, which determine uniquely the atomic structure of the corresponding rotated bilayer.

\begin{figure}[htbp]
\includegraphics*[width=\columnwidth]{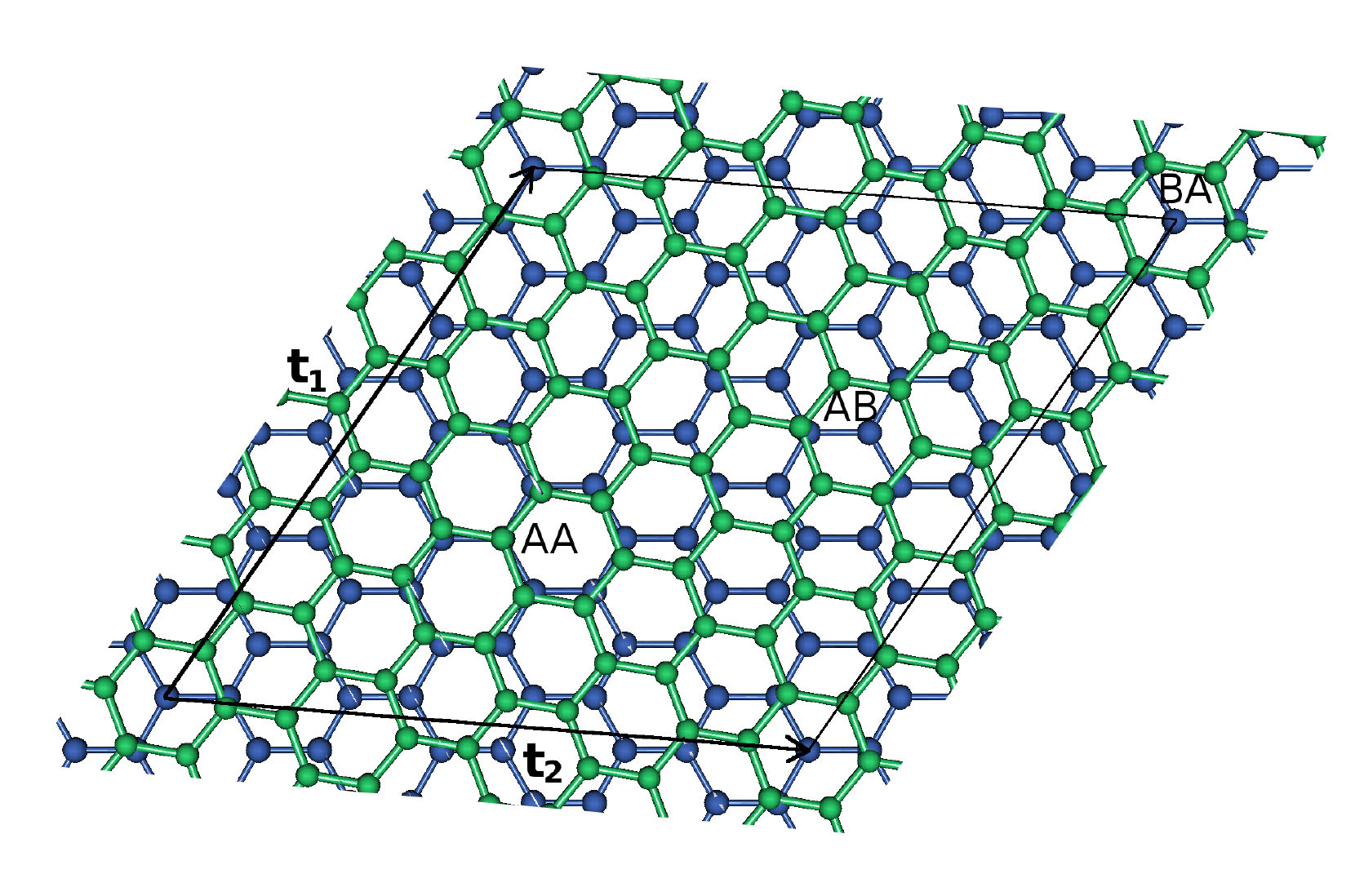}
\caption{A twisted unit cell (4,3) for a RRA of 9.43$^{\rm{o}}$; it contains 148 atoms. The AA, AB and BA sites are shown. 
 } \label{fig:uc}
\end{figure}

Fig. \ref{fig:uc} shows the unit cell of the $(4,3)$ bilayer, along with the unit cell vectors $ \vec{t}_{1}$ and $ \vec{t}_{2}$.
In a rotated bilayer, we distinguish four distinct regions, centered in specific sites given by the different possible atomic stackings. We label as AB 
a 
site with a top atom at the center of an hexagon. An AA site has an atom exactly on top of another, with a similar stacking for the nearest neighbors, and a BA site has an atom in the bottom layer exactly at the center of an hexagon of the top layer. These sites are indicated in Fig. \ref{fig:uc}.

The unit cells presenting distinct moir\'e patterns correspond to small angles (below 10$^{\rm o}$); therefore, in order to obtain large non-trivial unit cells, we choose \textit{n} and \textit{m} as two integers without any common divisors. By varying these numbers, any RRA from $0^{\rm o}$ to $60^{\rm o}$ can be attained.\cite{Campanera_2007}
It should be noted that structures with rather close RRA can be obtained from very different \textit{n} and \textit{m}. 
Such structures have similar moir\'e patterns, but they arise from unit cells with quite different number of atoms. 
We did most of our calculations for $m=n-1$. This choice gives the smallest unit cell for a set of RRA and covers fairly well the range of angles of interest.
However, other structures with dissimilar $n$, $m$ indices were generated in order to check that our results do not depend on the shape of the unit cell. 

For a bilayer graphene or graphite, moir\'e patterns are often observed for angles below $10^{\rm{o}}$. For instance, a (17,16),  with a $2^{\rm{o}}$ RRA, results in a unit cell with 3268 atoms.  
Setting the rotation axis on a B site and a unit cell with $m=n-1$ gives a unique AA site and an AB site 
 along the diagonal of the unit cell. The AA site lies at $1/3$ of the diagonal and the AB site lies at $2/3$. From our construction we get at every corner of our unit cell a BA site. There are also some other interesting points inside the unit cell; we will call a slip site (SL)\cite{Campanera_2007} a point like that in the middle of the line joining an AB site and the end of the main diagonal line. A slip site apparently looks as derived by a relative translation of the two layers, hence its name.
 
\subsection{Tight Binding Model}
\label{sec:TB}
We model the bilayer graphene band structure within the tight binding approximation. Within each layer, we consider a fixed nearest-neighbor intralayer interaction $\gamma_0=3.16$ eV. For the layer-layer interaction we consider a distance-dependent hopping between more than nearest layer-to-layer neighbors. 
 Thus, the hamiltonian is
given by $H=H_{1}+H_{2}+H_{int}$, where $H_{1}$ and $H_{2}$ are the hamiltonians for the top and the bottom layer 
and $H_{int}$ describes the interlayer coupling,
 \begin{equation}
H_{int} = - \sum_{i,j} \gamma_{1}e^{-\beta (\mathbf{r_{ij}}-\mathbf{d})} c^\dagger_{i} c_{j} + h.c.,
\label{eq:ham}
\end{equation}
where $\gamma_{1}=0.39$ eV is the nearest-neighbor interlayer hopping parameter, $\mathbf{d}$ is the interlayer distance, $\mathbf{r_{ij}}$ is the distance between atom $i$ on the top layer and atom $j$ on the bottom layer, and $\beta=3$. This value of $\beta$ reproduces accurately the dispersion bands calculated with a Density Functional Theory (DFT) approach.\cite{EricFlatBands,Shallcross_2008,Latil_2007} Besides, it gives a value of $\gamma_{3}$, the second nearest neighbor interlayer hopping, in agreement with values employed by other authors.\cite{Peeters2006,Chung2002} We let every atom in the top layer to interact with the atoms in the bottom layer located inside a circle of radius $3 a_0$, 
taking into account the complexity of the unit cell. This breaks the electron-hole (e-h) symmetry due to the fact that we are mixing the two sublattices. The asymmetry depends on the value of $\beta$ and on the RRA, being larger for smaller angles; it 
manifests in the difference between the electron and hole velocities.  
We find a larger velocity for electrons, as in the experimental results obtained by Luican {\it et al.}, \cite{Luican_2011} with a greater e-h asymmetry in the experimental values. 

In order to compute the charge density and total energy, we perform a sum over the Brillouin Zone (BZ); the $k$ points were selected with a Monkhorst-Pack scheme,\cite{Monkhorst_1976} taking into account the symmetries of the BZ. Convergency tests were made to select the appropriate number of $k$ points for the mesh. 

Under an external bias, the LDOS\cite{Campanera_2007} and the charge density in a twisted unit cell are inhomogeneous. This, along with the discreteness of the crystal lattice points, lead us to compute the Fourier transform (FT) of the charge excess density. As there are large oscillations of the charge density between the two sublattices, we computed separately the FT contributions of the A and B atoms.  
Furthermore, the FT allows us to evaluate in a continuous way the charge density on AB, BA, and AA regions on the top layer.
Only the zero order and the three lower components were needed to fully reproduce the charge distribution.


\section{Results and Discussion}
 \label{sec:results}

As discussed in the preceding section, when a potential difference is applied to a coupled bilayer, charge is transferred from one layer to the other. In the geometry chosen, there is an excess electronic charge $n$ that builds up on the top layer.  Note that if the layers are uncoupled, there is also a charge transfer due to the relative shift of the Dirac cones in the two layers, but the amount will be smaller than in a bilayer with nonzero coupling. 

\begin{figure}[htbp]
\includegraphics[width=\columnwidth]{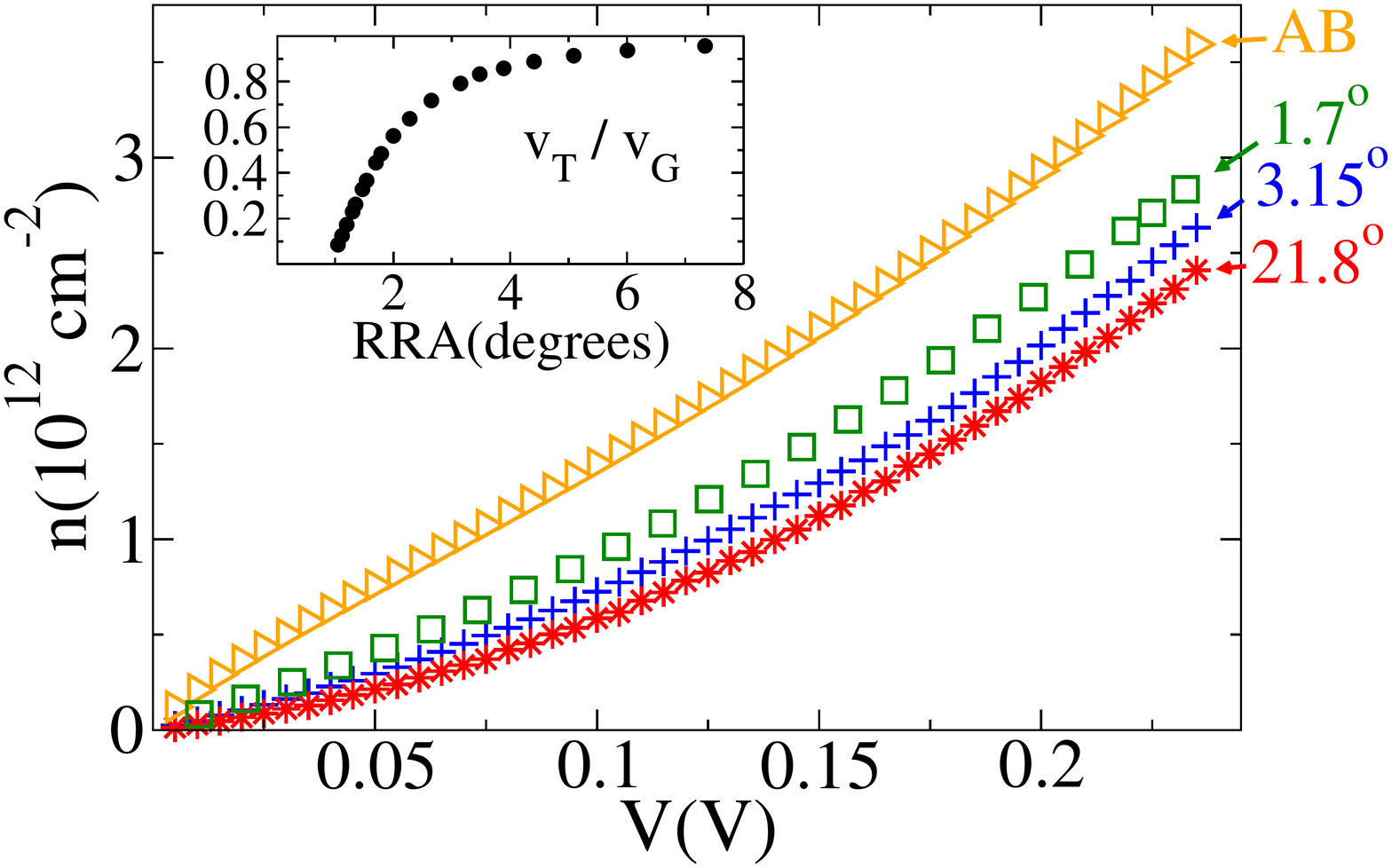}
\caption{(Color online) Excess electronic density on the top layer as a function of $V$ for several rotated bilayer graphene structures. The rotation angles are indicated in the Figure. Inset: Fermi velocity in twisted bilayer graphene, $v_T$, normalized with respect to that of monolayer graphene, $v_G$, as a function of the rotation angle RRA.
 } \label{fig:nvsV}
\end{figure}

In Fig. \ref{fig:nvsV} we plot the excess electronic density on the top layer as a function of $V$ for several rotated bilayers, including the AB case for comparison. We see how the excess electronic density diminishes with increasing rotation angle, indicating that the coupling between the rotated graphene layers is reduced for higher RRA, albeit non-negligible in any case. The values for large angles are rather similar; this saturation is correlated to the behavior of the carriers' velocity for increasing angle, which is equal to that of monolayer graphene for large RRA, see inset of Fig. \ref{fig:nvsV}. 

%
%

\begin{figure}[htbp]
\includegraphics*[width=\columnwidth]{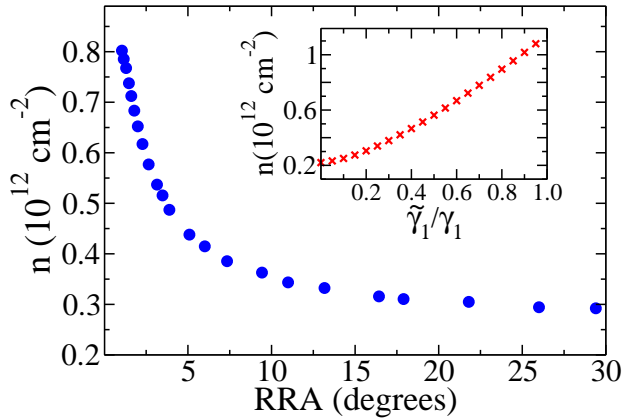}
\caption{(Color online) Excess electronic density on the top layer as a function of the rotation angle for a fixed electric field ($V=0.08$ V). Inset: Excess charge density as a function of the interlayer coupling for AB bilayer graphene. } \label{fig:cargavsangle}
\end{figure}

In  Fig. \ref{fig:cargavsangle} we show the angle dependence of the charge transfer for a fixed, nonzero electric field ($V=0.08$ V). The excess electronic charge has a strong variation for small angles, and then tends to a constant value, although there is a smooth variation even in the range from $20^{\rm o}$ to $ 30^{\rm o}$. However, this limiting value for the excess electronic charge is much higher than the one obtained for an uncoupled bilayer: the inset of Figure \ref{fig:cargavsangle} depicts the excess charge on a bilayer AB subject to the same electric field as a function of the interlayer coupling constant. We label this fictitious varying interlayer hopping 
$\tilde{\gamma_1}$ to distinguish it from the parameter ${\gamma_1}$ defined in Section \ref{sec:TB}. We see that the charge transferred between layers in an uncoupled system ($\tilde{\gamma_1}=0$) is $n=0.21 \times 10^{12} $ cm$^{-2}$, roughly 30\% smaller than the excess electronic charge for the largest RRA value in Fig.  \ref{fig:cargavsangle}, almost $n=0.3 \times 10^{12} $ cm$^{-2}$. 
Therefore, reaching a saturation in the charge transfer does not necessarily mean that the twisted bilayers are uncoupled. 
Notice as well that the charge with diminishing angle tends to the value of the perfect stacked AB bilayer.

\begin{figure}[thbp]
\includegraphics*[width= \columnwidth]{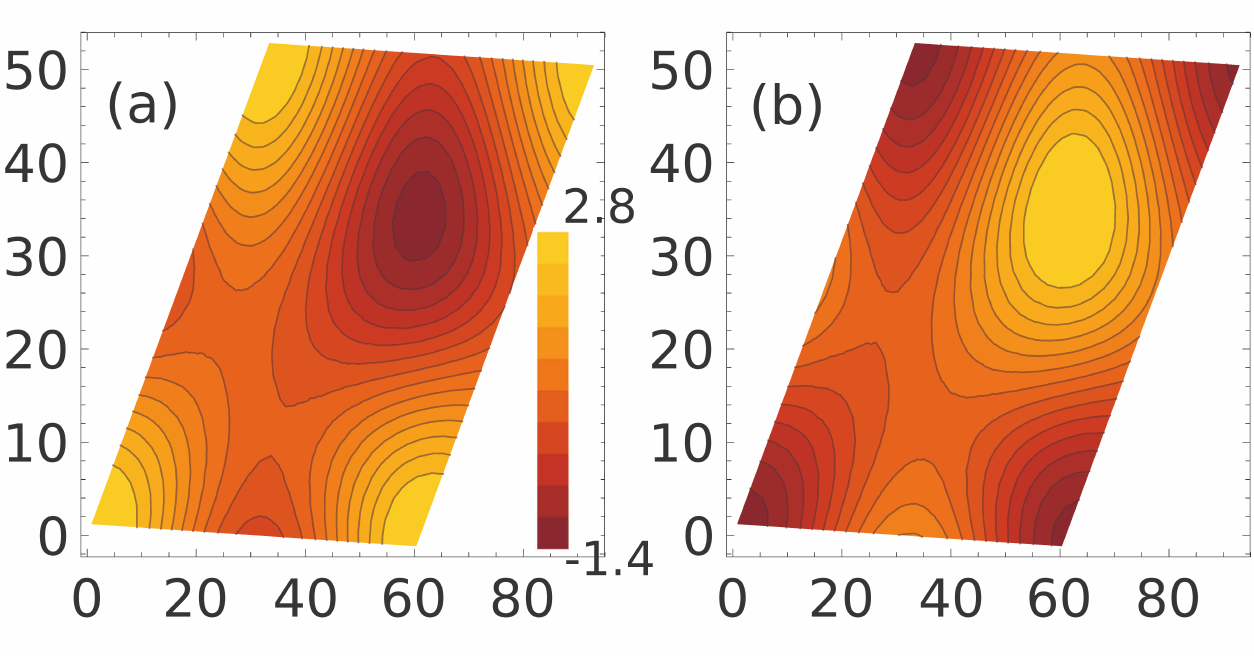}
\caption{(Color online) Contour plot of the excess electronic density, in units of $10^{12}$ cm$^{-2}$, on the top layer in the A (left) and B (right) atoms for the $(15,14)$ rotated bilayer, with 2524 atoms in the unit cell. The position coordinates ($x$ and $y$ axes) are in \AA .  
 } \label{Tn15m14}
\end{figure}

The excess charge presents an inhomogeneous spatial distribution that depends on the rotation angle. As there is a strong oscillation in the density between neighboring A and B sites, we choose to plot it for these points separately. In Fig. \ref{Tn15m14} we show the excess density on the top layer for a (15,14) bilayer. Different regions with respect to the electronic density distribution can be distinguished, corresponding to the stacking regions previously identified (see Fig. \ref{fig:uc}). For example, the BA zones, located at the corners of the unit cell, present an increase in the electronic density in the A atoms, whereas it decreases on the B atoms. On the other hand, AB zones display the opposite behavior: the A atoms show a reduction of the electronic charge, which in turn increases on the B atoms. The asymmetry is quite large, as can be seen 
in Fig. \ref{Tn15m14}. The excess charge density in B (A) atoms in AB (BA) sites is 300 \% larger than the averaged value ($n=0.7 \times 10^{12} $ cm$^{-2}$, see Fig. 
\ref{fig:cargavsangle}), varying from $-1.4$ to $2.8 \times 10^{12} $ cm$^{-2}$.  
On the other hand, the AA regions present smaller charge fluctuations, with similar electronic density values for A and B atoms. This can be understood by considering the local surroundings of the atoms: the AB or BA regions resemble the stacked AB bilayer, where there is a difference between the LDOS and the charge density in both sites, whereas the AA region is more like the stacked AA bilayer, with similar density values in A and B sites.

%



\begin{figure}[htbp]
\includegraphics*[width=\columnwidth]{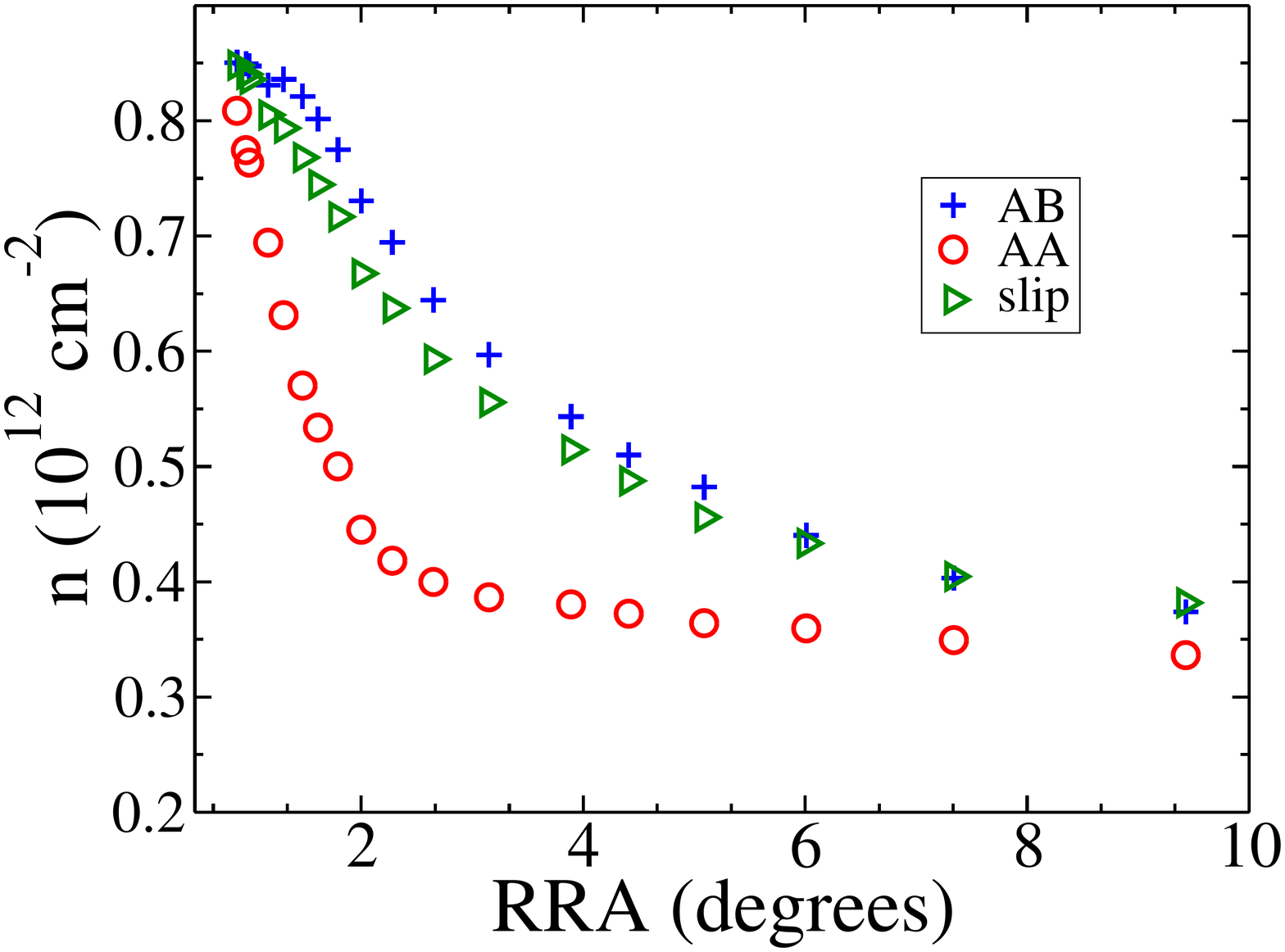}
\caption{(Color online) Excess electronic density on the top layer as a function of the rotation angle for a fixed potential $V=0.08$ V for different stacking regions on the unit cell, namely, AA, AB and slip. 
 } \label{fig:carga_regions}
\end{figure}

Given such differences, we have studied the charge distribution in different stacking regions as a function of the rotation angle in the top layer. 
We average the FT of the excess charge density in atoms A and B, evaluated at the AA, AB, BA, and slip points of the unit cell. 
Fig. \ref{fig:carga_regions} shows this averaged FT for a fixed $V=0.08$ V at the AA, AB and slip regions for angles below $10^{\rm o}$. As there is a symmetry under the interchange of A and B sites on the AB and BA regions, only the values of the AB site are shown. Several features should be noted: (i) the AA site takes less charge than the AB site; (ii) the slip site density behavior is very similar to that of the total density;  and  (iii) for very low angles, the density is more homogeneous, as opposed to the LDOS, which peaks at AA sites near the Fermi energy for small RRA.\cite{Trambly_2010}


\begin{figure}[htbp]
\includegraphics*[clip,width=\columnwidth]{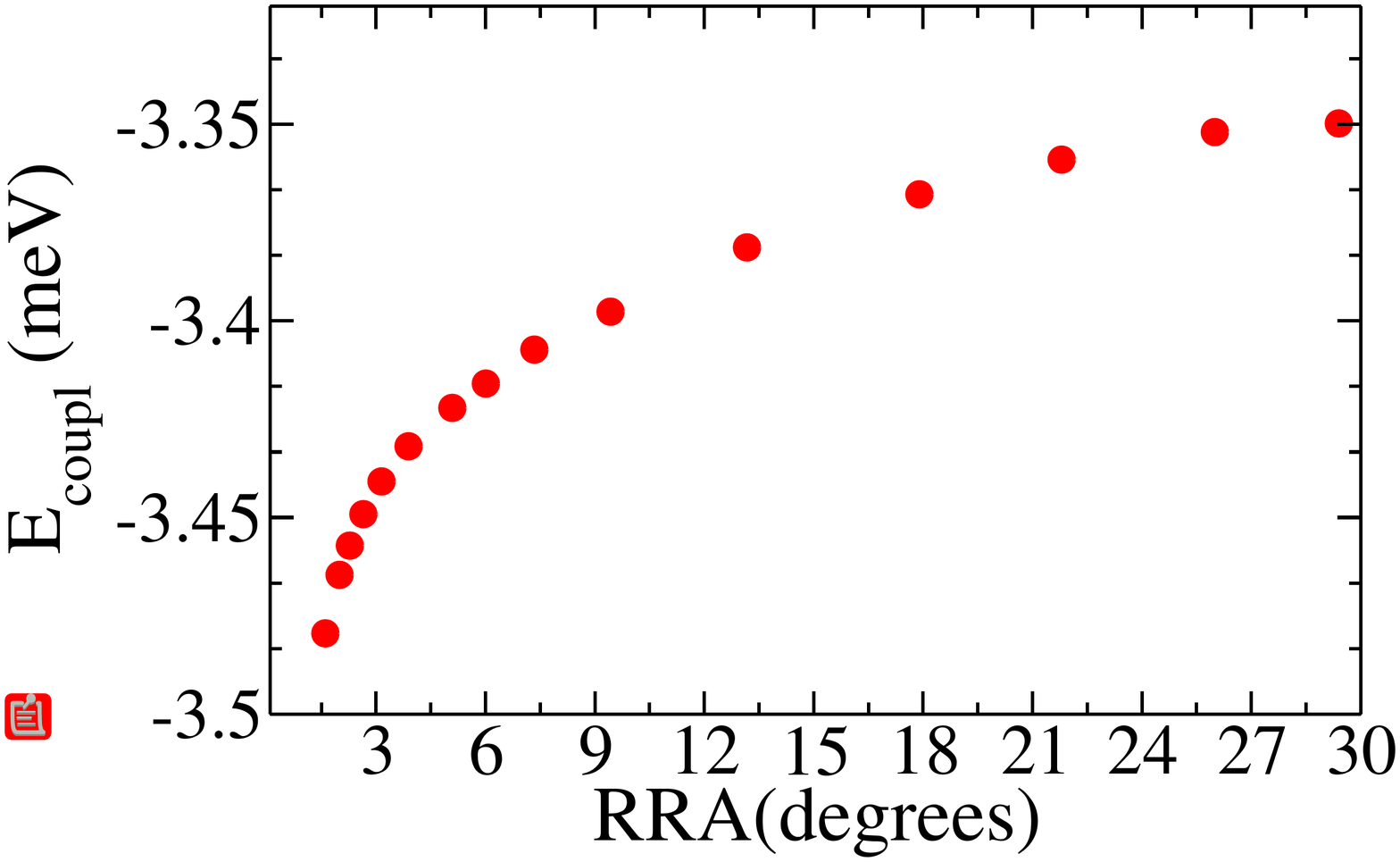}
\caption{(Color online) Total coupling energy per atom of the twisted bilayers as a function of the RRA. 
 } \label{fig:energ}
\end{figure}

Besides these magnitudes related to the charge transfer under an applied perpendicular electric field, one way to theoretically assess the electronic coupling is to calculate the interlayer coupling kinetic energy in the unbiased double layer. We define $E_{coupl}$ as the difference between the total kinetic energy of the coupled bilayer and the kinetic energy of the uncoupled double layer, $ E_{coupl} =\langle \psi | H_{int} | \psi  \rangle $, where $| \psi \rangle$ is the ground state wavefunction of the total coupled bilayer hamiltonian. This quantity is directly related to the interlayer interaction, which should vary with the RRA. Fig. \ref{fig:energ} presents the interlayer coupling energy $E_{coupl}$ as a function of the rotation angle. We see that although $E_{coupl}$ increases with increasing rotation angle, is by no means negligible for RRA over 20$^{\rm o}$.
The energy for low angles decreases, tending to the value of the stacked AB bilayer, $-5.1$ meV.

Finally, we note that in the analysis of the bias-induced charge transfer we have neglected the Coulomb energy $E_C$ corresponding to the two charged layers.\cite{McCann_2006b,FRossier_2007} We can estimate the charging energy $E_C$ by considering that the bilayer is like a capacitor with oppositely charged plates, and assuming that the electronic density is uniformly distributed in the two-dimensional layers. With these assumptions, 
$E_C= \frac{e^2 n^2 S}{2 \epsilon} d_0$, 
where $e$ is the elementary charge, $d_0$ is the interlayer separation, $S$ is the sample area, and $\epsilon$ is the dielectric constant of the medium. This Coulomb energy induces an extra potential between the layers given by $\Delta V(n) = \frac{e^2 n d_0}{\epsilon}$. Because of the small separation between the two graphene  layers, the Coulomb potential  induced by the charge transfer is much smaller than the externally applied bias, so therefore it is justified to neglect it in the calculation of the bias-induced charge imbalance.

\section{Conclusions}

\label{sec:concl}

In summary, we have shown that the coupling between rotated graphene bilayers can be evidenced by several means. Firstly, the application of an external electric field induces a charge transfer between layers that can be explored as a way to indicate the electronic coupling, this charge transfer depends on the RRA. Secondly, the spatial charge inhomogeneities are also a signal of the electronic coupling between layers, analogously to the differences in the LDOS reported by other authors.\cite{Campanera_2007} The coupling was theoretically estimated by computing the coupling energy, i.e., the difference in the energy of a coupled bilayer and the energy of two uncoupled graphene layers. This magnitude is nonzero for all angles, thus evidencing that rotated graphene bilayers are coupled for all rotation angles. The values of the excess charge density and the coupling energy tends toward the stacked AB bilayer for low twist angles.

\begin{acknowledgments}
This work has been partially supported by MEC-Spain under grant
FIS2009-08744 and by the CSIC/CONICYT program, grant 2009CL0054.
E.S.M. acknowledges CONICYT (Chile) and UTFSM for the internal grant PIC-DGIP. P.V. acknowledges FONDECYT grant 1100508. 
\end{acknowledgments} 


\end{document}